\documentclass{PoS}

\usepackage{amsmath}
\usepackage{booktabs}

\rightline{DO-TH 17/25, QFET-2017-17, TTP17-040}

\title{Rare radiative charm decays in the standard model and beyond}

\ShortTitle{Rare radiative charm decays in the standard model and beyond}

\author{\speaker{Stefan de Boer}%
       \\
       Institute for Theoretical Particle Physics (TTP), Karlsruhe Institute of Technology (KIT)\\
       Technical University of Dortmund\\
       E-mail: \email{stefan.boer@kit.edu}}

\abstract{

Motivated by the recent measurement of $D^0\to\rho^0\gamma$ improved standard model (SM) predictions for branching ratios and CP asymmetries of radiative charm decay are given.
Weak annihilation induced decays are probes of non-perturbative QCD approaches.
Rare decays probe the SM and physics beyond the SM, e.g.~leptoquark and supersymmetric models.
Opportunities with $\Lambda_c\to p\gamma$ for future polarization measurements are presented.

}

\FullConference{The European Physical Society Conference on High Energy Physics\\
         5-12 July\\
         Venice, Italy}

\begin{document}

\section{Introduction}

Flavor changing neutral current (FCNC) decays involving charm quarks, i.e.~$c\to u$ transitions, are rare in the standard model (SM) since they are loop suppressed.
Furthermore, the GIM mechanism sets, e.g., CP asymmetries to approximate zero in the SM.
On the other hand, branching ratios are controlled by non-perturbative contributions.
Hence, rare charm decays can be used, first, as a probe of physics beyond the standard model (BSM) and, second, to probe models and frameworks which are employed to calculate non-perturbative SM contributions.

Motivated by the recent, first measurement of the branching ratio and CP asymmetry of the decay $D^0\to\rho^0\gamma$ \cite{Abdesselam:2016yvr},
\begin{align}\label{eq:D0rho0gamma_Belle}
 &\mathcal B(D^0\to\rho^0\gamma)=(1.77\pm0.30\pm0.07)\times10^{-5}\,,&&A_{CP}=0.056\pm0.152\pm0.006\,,
\end{align}
we study rare and weak annihilation (WA) induced radiative decays.
This study is based on \cite{deBoer:2017que}.
The perturbative inclusive SM prediction $\mathcal B(D\to X_u\gamma)=\mathcal O(10^{-8})$ \cite{Greub:1996wn} is too low compared with the exclusive measurement (\ref{eq:D0rho0gamma_Belle}), pointing towards large power corrections, resonant contributions and/or BSM physics.

We give predictions for radiative decays of $D$ mesons into a vector meson in the SM, in leptoquark (LQ) and supersymmetric (SUSY) models in section \ref{sec:DtoVgamma}.
Section \ref{sec:Lambda_ctopgamma} is on baryonic $\Lambda_c\to p\gamma$ decays and their opportunities for future polarization measurements.
A summary is given in section \ref{sec:summary}.

\section{The decay $D\to V\gamma$}
\label{sec:DtoVgamma}

Generically, the effective weak Lagrangian is
\begin{align}
 \mathcal L_\text{eff}(\mu\sim m_c)\sim\sum_iC_iQ_i
\end{align}
with the operators
\begin{align}
 &Q_{2(1)}=(\bar u_L\gamma_\mu(T^a)q_L)(\overline q_L\gamma^\mu(T^a)c_L)\,,\nonumber\\
 &Q_7^{(\prime)}=\frac{e\,m_c}{16\pi^2}(\bar u_{L(R)}\sigma^{\mu\nu}c_{R(L)})F_{\mu\nu}\,,\nonumber\\
 &Q_8^{(\prime)}=\frac{g_s\,m_c}{16\pi^2}(\bar u_{L(R)}\sigma^{\mu\nu}T^ac_{R(L)})G^a_{\mu\nu}\,,
\end{align}
see \cite{deBoer:2017que} for details.
The SM (effective) Wilson coefficients $C_i^\text{(eff)}$ are known to two loop in QCD \cite{deBoer:2017que,Greub:1996wn,Fajfer:2002gp,deBoer:2016dcg}.

\begin{figure}
 \centering
 \includegraphics[trim=3cm 23.5cm 7cm 4cm,clip,width=0.8\textwidth]{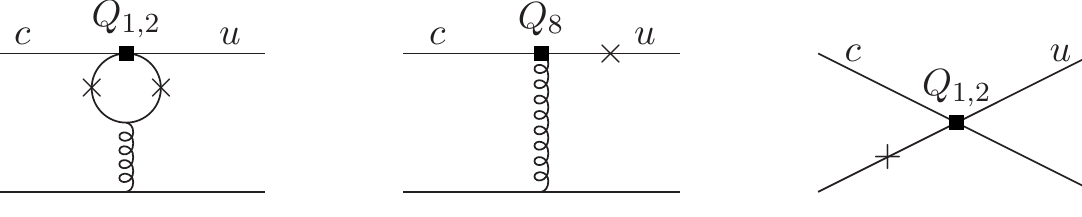}
 \caption{Hard spectator interaction (first two) and WA (third) diagrams.
 The crosses indicate photon emission.
 Figure adopted from \cite{deBoer:2017que}.}
 \label{fig:diagrams}
\end{figure}

To include corrections to the perturbative Wilson coefficients we employ two frameworks: \textit{(1)} a QCD based approach, worked out for $b$ physics in \cite{Bosch:2001gv,Bosch:2004nd}, and \textit{(2)} a hybrid model of the heavy quark effective theory and chiral perturbation theory using experimentally measured parameters \cite{Fajfer:1997bh,Fajfer:1998dv}.
In the first approach, we compute the leading power corrections $\sim\tfrac1{m_D}$, shown in figure~\ref{fig:diagrams}.
They involve the spectator quark and depend on $\lambda_D$, the first negative moment of the $D$ meson light-cone distribution amplitude.
The parameter $\lambda_D\sim\mathcal O(0.1\,\text{GeV})$ can presently only be estimated.

\begin{figure}
 \centering
 \includegraphics[width=0.45\textwidth]{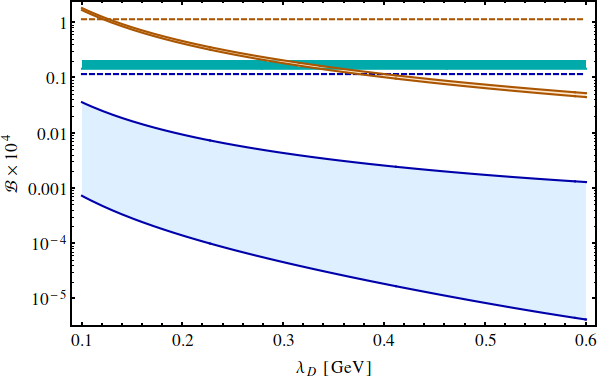}
 \hfil
 \includegraphics[width=0.45\textwidth]{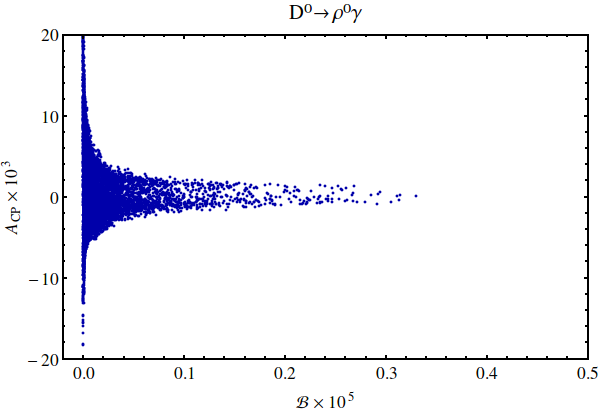}
 \caption{\textit{Left:} Branching ratios of $D\to\rho\gamma$ in the SM as a function of $\lambda_D$.
 The upper orange curves are for $D^+\to\rho^+\gamma$ and the lower blue curves show $D^0\to\rho^0\gamma$.
 The solid curves/bands represent approach (1), dashed lines are the maximal predictions in approach (2) and the cyan band depicts the measured branching ratio of $D^0\to\rho^0\gamma$ \cite{Abdesselam:2016yvr}.
 \textit{Right:} CP asymmetries versus branching ratios for $D^0\to\rho^0\gamma$ in the SM approach (1).
 The measured CP asymmetry at one $\sigma$ (\ref{eq:D0rho0gamma_Belle}) covers the shown range, whereas the measured one $\sigma$ branching ratio is above it.
 Figures adopted from \cite{deBoer:2017que}.}
 \label{fig:BlambdaDDrhogamma_ACPB}
\end{figure}

The branching ratios of $D\to\rho\gamma$, as predicted in both SM approaches, are shown in figure~\ref{fig:BlambdaDDrhogamma_ACPB}.
The parameter $\lambda_D$ can be constrained by measuring the branching ratio of the charged decay $D^+\to\rho^+\gamma$.
The branching ratio of the neutral decay $D^0\to\rho^0\gamma$ in approach (1) is subject to larger uncertainties due to the color suppressed combination of Wilson coefficients.
In approach (2), the predictions cover predictions from other approaches \cite{Burdman:1995te,Khodjamirian:1995uc}, see \cite{deBoer:2017que}.
Compared to the measured $D^0\to\rho^0\gamma$ branching ratio (\ref{eq:D0rho0gamma_Belle}), predictions in both approaches are too low.
This may be addressed to unknown corrections.
However, a (B)SM interpretation of data is feasible for CP asymmetries.
For $D^0\to\rho^0\gamma$, the $|A_{CP}^\text{SM}|<2\times10^{-2}$ if the measured branching ratio is explained by the SM, see figure~\ref{fig:BlambdaDDrhogamma_ACPB}.
Since the uncertainties of the present experimental data are controlled by statistics future experiments will test the SM.

The approaches (1) and (2) can be probed with WA induced radiative decays of neutral $D$ mesons.
In table~\ref{tab:Dphigamma_DKstar0gamma_branching_ratios} predictions and data are given for the branching ratios of $D^0\to(\phi,\bar{K^*}^0)\gamma$ decays.
\begin{table}
 \centering
 \begin{tabular}{crr}
  \toprule
  branching ratio                 &  $D^0\to\phi\gamma$            &  $D^0\to\bar{K^*}^0\gamma$  \\
  \midrule
  approach (1)                    &  $(0.0074-1.2)\times10^{-5}$   &  $(0.011-1.6)\times10^{-4}$  \\
  approach (2)                    &  $(0.24-2.8)\times10^{-5}$     &  $(0.26-4.6)\times10^{-4}$  \\
  \midrule
  data \cite{Abdesselam:2016yvr}  &  $(2.76\pm0.21)\times10^{-5}$  &  $(4.66\pm0.30)\times10^{-4}$  \\
  data \cite{Aubert:2008ai}       &  $(2.81\pm0.41)\times10^{-5}$  &  $(3.31\pm0.34)\times10^{-4}$  \\
  \bottomrule
 \end{tabular}
 \caption{Branching ratios in the SM approach (1) scale as $(\tfrac{0.1\,\text{GeV}}{\lambda_D})^2$.
 Table adopted from \cite{deBoer:2017que}.}
 \label{tab:Dphigamma_DKstar0gamma_branching_ratios}
\end{table}
The data and the SM predictions are consistent.
However, a slow convergence of the expansions in $\tfrac1{m_D}$, $\alpha_s$ is indicated.

As two BSM scenarios, we study scalar and vector LQ models, see \cite{deBoer:2017que}, and SUSY models for which we use the mass insertion approximation \cite{Gabbiani:1996hi}.
In both models, the Wilson coefficients $C_{7,8}^{(\prime)}$ are loop induced, correlated and constrained by other observables \cite{deBoer:2017que,deBoer:2016dcg}.
The induced branching ratios and CP asymmetries are given in table~\ref{fig:LQ_SUSY}.
\begin{table}
 \centering
 \begin{tabular}{ccc}
  \toprule
  model  &  branching ratio           &  CP asymmetry                \\
  \midrule
  LQ     &  SM-like                   &  $\lesssim\mathcal O(10\%)$  \\
  SUSY   &  $\lesssim2\times10^{-5}$  &  $\lesssim0.2$               \\
  \bottomrule
 \end{tabular}
 \caption{Branching ratio and CP asymmetry of $D^0\to\rho^0\gamma$.}
 \label{fig:LQ_SUSY}
\end{table}
Generically, contributions from SUSY models can be larger than LQ model contributions since the SUSY Wilson coefficients are enhanced by the gluino mass.
Specifically, LQs can induce $A_{CP}^\text{LQ}\lesssim\mathcal O(10\%)$, whereas SUSY induced observables can be close to the experimental data (\ref{eq:D0rho0gamma_Belle}).

\section{The decay $\Lambda_c\to p\gamma$}
\label{sec:Lambda_ctopgamma}

\begin{figure}
 \centering
 \includegraphics[width=0.7\textwidth]{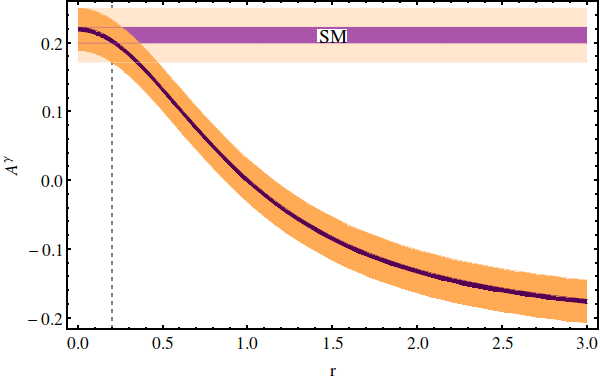}
 \caption{Asymmetry $A^\gamma$ for $P_{\Lambda_c}=-0.44$, approaching $\tfrac{P_{\Lambda_c}}2$ as $r\to\infty$.
 The bands represent the statistical uncertainties for $N=10^3$ (orange) and $N=10^5$ (purple).
 In the SM and LQ models $A^\gamma(r\lesssim0.2)$, indicated by the dashed line, and the corresponding $A^\gamma$ ranges are shown by the horizontal bands.
 In SUSY $r\lesssim\mathcal O(1)$.
 Figure adopted from \cite{deBoer:2017que}.}
 \label{fig:AgammaZ}
\end{figure}

We infer the $\Lambda_c\to p\gamma$ branching ratio to be \cite{deBoer:2017que}
\begin{align}
 \mathcal B_{\Lambda_c\to p\gamma}\sim\mathcal O(10^{-5})\,,
\end{align}
hence, the expected number $N$ of decays at future colliders as given in table~\ref{tab:NLambdacpgamma}.
\begin{table}
 \centering
 \begin{tabular}{ccc}
  \toprule
  collider  &  number             &  reference  \\
  \midrule
  Belle II  &  $\sim[10^3,10^4]$  &  $L\simeq5\,\text{ab}^{-1}$ \cite{Aushev:2010bq}  \\
  FCC-ee    &  $\sim10^5$         &  $N(Z)\sim10^{12}$ \cite{dEnterria:2016fpc}  \\
  \bottomrule
 \end{tabular}
 \caption{Expected number of $\Lambda_c\to p\gamma$ decays within one year, neglecting reconstruction efficiencies.}
 \label{tab:NLambdacpgamma}
\end{table}
Baryonic $\Lambda_c$ decays induce an additional observable, the forward-backward asymmetry of photon momentum relative to $\Lambda_c$ boost, worked out for $b$ physics in \cite{Hiller:2001zj}.
The angular asymmetry
\begin{align}
 &A^\gamma=-\frac{P_{\Lambda_c}}2\frac{1-|r|^2}{1+|r|^2}\,,&&r=\frac{C_7^\prime}{C_7}
\end{align}
probes the handedness of $c\to u\gamma$ transitions.
Here, the $\Lambda_c$ polarization, inherited from the decay of a $Z$ boson into a charm-anticharm quark pair,
\begin{align}
 P_{\Lambda_c}^{(Z)}\simeq-0.44\pm0.02\,,
\end{align}
where the parametrization of \cite{Falk:1993rf,Galanti:2015pqa,Kats:2015zth} is used.
The polarization is measurable at future colliders as well as BaBar, Belle and LHC.
The asymmetry $A^\gamma$ in the SM, LQ and SUSY models is shown in figure~\ref{fig:AgammaZ}.
In SUSY models, $A^\gamma$ can be different from the SM prediction, including a sign flip, pointing out an opportunity for future polarization measurements with baryons.

\section{Summary}
\label{sec:summary}

Motivated by the first measurement of a rare radiative charm decay, we have presented a work of $D\to V\gamma$ and $\Lambda_c\to p\gamma$ \cite{deBoer:2017que}.
For $D\to V\gamma$, we have given improved SM predictions including power corrections and updating a hybrid model.
Predictions for branching ratios are uncertain, whereas CP asymmetries are approximate SM null tests.
Branching ratios, however, test non-perturbative QCD approaches, once further rare and weak annihilation induced radiative decays of $D$ mesons are measured.
On the other hand, measurements of CP asymmetries will constrain or reveal BSM physics, e.g.~LQ and SUSY models.
For $\Lambda_c\to p\gamma$, we have explored opportunities with an additional angular observable for future colliders, e.g.~an FCC-ee.

\section*{Acknowledgements}

I thank the organisers for the wonderful conference.
I am grateful to Gudrun Hiller for a very enjoyable collaboration and Ivan Nisandzic for reading the manuscript.
This project is in part supported by the DFG Research Unit FOR 1873 ``Quark Flavour Physics and Effective Field Theories''.

\end{document}